\documentclass[aps,prd,
floatfix,preprintnumbers,superscriptaddress,nofootinbib,nameinlink,capitalise]{revtex4-2}

\usepackage[left=2.54cm, right=2.54cm, top=2.54cm, bottom=2.54cm]{geometry}
\usepackage{palatino}
\usepackage{amsmath, amssymb}
\usepackage{graphicx}
\usepackage{array, booktabs, multirow}
\usepackage{float}
\usepackage{float} 
\usepackage{hyperref}
\usepackage{subfig}
\usepackage{nameref}      
\usepackage{cleveref}
\usepackage{placeins}
\usepackage{verbatim}
\usepackage{tikz}
\usetikzlibrary{shapes.geometric, arrows}

\tikzstyle{startstop} = [rectangle, rounded corners, minimum width=3cm, minimum height=1cm,text centered, draw=black, fill=red!30]
\tikzstyle{process} = [rectangle, minimum width=3cm, minimum height=1cm, text centered, draw=black, fill=blue!20]
\tikzstyle{decision} = [diamond, minimum width=3cm, minimum height=1cm, text centered, draw=black, fill=green!30]
\tikzstyle{arrow} = [thick,->,>=stealth]

\usepackage{textgreek}
\usepackage[justification=raggedright,singlelinecheck=false]{caption}
\usepackage{color}
\newcommand{\bmat}{\left(\begin{array}}
\newcommand{\emat}{\end{array}\right)}
\newcommand{\be}{\begin{equation}}
\newcommand{\ee}{\end{equation}}
\newcommand{\bea}{\begin{eqnarray}}
\newcommand{\eea}{\end{eqnarray}}

\usepackage{graphicx}
\usepackage{subcaption}

\begin{document}

\title{Hybrid Ensemble Method for Detecting Cyber-Attacks in Water Distribution Systems Using the BATADAL Dataset}

\author{Waqas Ahmed}
\email{waqasmit@hbpu.edu.cn}
\affiliation{Center for Fundamental Physics, School of Artificial Intelligence, Hubei Polytechnic University, Huangshi 435003, China}
\affiliation{The European Higher Education Institute,
St. Julian's STJ 3141, Malta}



\noaffiliation


\vspace{1em}
\begin{abstract}

The cybersecurity of Industrial Control Systems that manage critical infrastructure such as Water Distribution Systems has become increasingly important as digital connectivity expands. BATADAL benchmark data is a good source of testing intrusion detection techniques, but it presents several important problems, such as imbalance in the number of classes, multivariate time dependence, and stealthy attacks. We consider a hybrid ensemble learning model that will enhance the detection ability of cyber-attacks in WDS by using the complementary capabilities of machine learning and deep learning models. Three base learners, namely, Random Forest , eXtreme Gradient Boosting , and Long Short-Term Memory  network, have been strictly compared and seven ensemble types using simple averaged and stacked learning with a logistic regression meta-learner. Random Forest analysis identified top predictors turned into temporal and statistical features, and Synthetic Minority Oversampling Technique (SMOTE) was used to overcome the class imbalance issue. The analyics indicates that the single Long Short-Term Memory  network model is of poor performance (F1  = 0.000, AUC = 0.4460), but tree-based models, especially eXtreme Gradient Boosting, perform well (F1  = 0.7470, AUC=0.9684). The hybrid stacked ensemble of Random Forest , eXtreme Gradient Boosting , and Long Short-Term Memory  network  scored the highest, with the attack class of 0.7205 with an F1-score and a AUC of 0.9826 indicating that the heterogeneous stacking between model precision and generalization can work. The proposed framework establishes a robust and scalable solution for cyber-attack detection in time-dependent industrial systems, integrating temporal learning and ensemble diversity to support the secure operation of critical infrastructure.

\end{abstract}

\maketitle

\newpage
\section{Introduction}\label{sec:introduction}

The networks that constitute the operational backbone of critical infrastructure are known as Industrial Control Systems (ICS) and Supervisory Control and Data Acquisition (SCADA) \cite{ref1,cisa_ics}. These systems control crucial services, including, including water supply, electricity production, and transportation sources. Increasing the interplay between Operational Technology (OT) and Information Technology (IT) has increased efficiency but also increased the cyber attack surface significantly \cite{wef2023, gartner}. This in turn has made ICS environments high-ranking targets of various malicious attackers including individual members of hackers as well as state-sponsored entities \cite{sans}. Such attacks as the 2021 Colonial Pipeline ransomware attack reflect the devastating effects of such attacks in real life with damage to the economy, the environment, and even human lives being in danger of danger \cite{cisa}.

The Water Distribution System (WDS) is one of the most sensitive and essential parts. An effective cyber-physical attack may cause disinfection of the chemical dosage by altering the work of pumps, harming the supply through the manipulation of tanks, and causing crises in population health and environmental harm \cite{taormina2017characterizing,cisa2021water}. In order to promote innovation in the development of advanced detection algorithms to WDS, the BATADAL (Battle of the Attack Detection Algorithms) competition offered an open and realistic dataset to represent the essence of the challenges involved in the security of ICS \cite{wa1}.

Despite the serious progress already achieved in using machine learning to cybersecurity to identify intrusions, several unsolved issues can and do exist in the ICS field, many of which, as illustrated in the BATADAL dataset, can be viewed as examples of this case. One of the major problems is the issue of the high imbalance of classes. Cyber-attacks do not occur that often, but usually, they occur in a chain of ordinary operational data. It follows that this data becomes disproportionately high and the models can easily obtain artificially high accuracy by merely predicting the majority class (normal operation) and keeping failing to predict the minority (attacks) when attempting to predict it \cite{nankya2020addressing}. The result of an imbalance is that it produces large false-negatives, and fails to detect critical threats.The other fundamental problem is that ICS data is multivariate and time-series. Dangerous actions are a gradual process that is usually multi-stage, and not an isolated exception. Classical models of machine learning that presuppose independent and identically distributed (i.i.d.) data samples do not support such crucial temporal dependencies \cite{kim2021multivariate}. Moreover, intruders can follow stealthy approaches to mask their operations as noise during the normal functioning, which makes it even harder to detect the attack on the noise of a normal operation \cite{urbina2016limiting}. 

There is a further dilemma of model selection, there are algorithms with differing inductive biases and representation strengths. Random Forest (RF) \cite{breiman2001random}and XGBoost (XGB) \cite{chen2016xgboost}can be employed as tree-based learners that are highly beneficial in revealing nonlinear interactions in static tabular data, and LSTM networks \cite{hochreiter1997long} can be used as recurrent learners that take advantage of sequential interactions in time-series signals.Dependability on any of the models can result in the unnoticed vulnerabilities as long as the attack patterns are outside the scope of the model.

Although the individual components of these algorithmic families have been studied previously\cite{gendreau2022,chandy2019} there is a strong gap in the systematic construction and empirical validation of a hybrid framework fusing these two complementary paradigms into one and strong systemic detection method used on Water Distribution Systems (WDS)\cite{housh2018}.

In this work makes several key contributions to the field of cyber-attack detection in Water Distribution Systems (WDS):

\begin{enumerate}
    \item \textbf{Comprehensive Data Exploration:} A detailed Exploratory Data Analysis (EDA) of the BATADAL dataset is presented, revealing its temporal behavior, variable interactions, and distinct attack signatures that characterize abnormal system operations.
    
    \item \textbf{Hybrid Stacked Ensemble Framework:} A novel stacked-ensemble architecture is proposed, combining the predictive strengths of Random Forest (RF), XGBoost (XGB), and Long Short-Term Memory (LSTM) networks through a logistic regression meta-learner.
    
    \item \textbf{Empirical Performance Validation:} The proposed hybrid approach achieves state-of-the-art performance on the BATADAL benchmark, consistently outperforming individual base models and simpler pairwise ensembles across multiple evaluation metrics.
    
    \item \textbf{Model Explainability via SHAP:} Explainable AI techniques, particularly SHAP value analysis, are employed to identify dominant temporal and operational features that drive detection outcomes, thereby improving interpretability and supporting operational trust.
    
    \item \textbf{Analysis of Model Complementarity:} A systematic discussion is provided to illustrate how the differing inductive biases of tree-based and sequential learners contribute to enhanced robustness, explaining why their integration yields superior detection accuracy compared to standalone models.
\end{enumerate}

The structure of the paper is organized as follows. In Section~\ref{sec:introduction}, we provide an introduction to the problem of cyber-attack detection in Water Distribution Systems (WDS) and highlight the challenges associated with time-dependent industrial control data. Section~\ref{sec:literature_review} presents a comprehensive review of the existing literature on ICS security, machine learning approaches, and hybrid ensemble methods. In Section~\ref{sec:methodology}, the dataset, preprocessing steps, feature engineering, and the proposed hybrid stacked ensemble framework combining Random Forest, XGBoost, and LSTM models are described in detail. Section~\ref{sec:results} reports the performance of individual base learners and hybrid ensembles, along with model interpretability using SHAP analysis. Section~\ref{sec:discussion} provides a critical discussion of the results, highlighting the advantages of the proposed approach and its implications for real-world cyber-attack detection. Finally, Section~\ref{sec:conclusion} concludes the paper and outlines potential directions for future research.

\section{Literature Review} \label{sec:literature_review}
As the field of ICS security has progressed,  \cite{scarfone2007guide}, The ICS security development has seen the advent of more advanced anomaly based detection systems, and the old signature-driven systems no longer prove to be effective in combating the emerging or polymorphic cyber threats as they appear today .  The first types of approaches used to detect anomalies were based on statistical process control models\cite{scarfone2007guide} and physical model-based residuals \cite{venkatasubramanian2003review}. Although these methods were theoretically sound, they tended to produce unreasonable false alarms in practice, and required substantial experience to use well in practice \cite{mitchell2014survey}.

Machine learning had brought revolutionary features in this field. Unsupervised methods such as Principal Component Analysis (PCA) \cite{greenacre2022principal}, K-Means clustering ( \cite{hartigan1979algorithm} and One-Class Support Vector Machines (OC-SVM) \cite{platt2001estimating} made possible the building of normative behavioral models with which operational deviations were detected. For instance, \cite{goh2016dataset} applied OC-SVM on the SWaT testbed data, demonstrating its utility when labeled attack data is scarce. However, supervised learning schemes tend to do much better than unsupervised strategies in most cases with enough accessible labeled information \cite{chandola2009anomaly}, particularly when the false alarm associated with anomalies is of operational importance such as with ICS usage where false alarms are important to operation health aspects of ICS (and not the quality of an operation) \cite{axelsson2000base, garcia2009anomaly}.

As the release of labeled ICS datasets  e.g. SWaT, WADI and BATADAL \cite{goh2016dataset, ahmed2017wadi, taormina2017characterizing,cisa2021water} where labels are assigned, supervised learning took center stage. Among them the tree-based ensemble techniques have always performed better. Random Forest (RF) which is a bunch of decision trees constructed using the bagging technique is well known by virtue of its ability to combat the problem of overfitting as well as its ability to handle high-dimensional data sets \cite{breiman2001random}. It featured in many BATADAL data analysis [28].

Sequential, machine-based techniques such as Gradient Boosting Machines (GBMs), which use trees to address the error committed by earlier trees, have also put performance limits even less soundly. Specifically, XGBoost (XGB) has gained great popularity as a standard of decision science competitions because it has proved to be computer efficient, regularized and state of the art with structured data \cite{chen2016xgboost}. It has been examined intensively and proven to be efficient in the utilization of cybersecurity, like network intrusion detection \cite{attia2020network}. The synergy of such tree-based models is that by being highly non-linear in their interactions between features at a given time they can produce powerful snapshot feature analyzers.
.

Time-varying nature of ICS sensors makes a highly valid point to consider time-series-based deep learning structures. It is possible to also train RNNs and more importantly Long Short term Memory (LSTM) networks, which address the problem of vanishing gradients of vanilla RNNs, to learn long time temporal dependencies \cite{hochreiter1997long}. This makes them conceptually well-posed to determine multi-step attack campaigns, which can have hours or days to develop. LSTMs were employed by \cite{dash2025optimized} to use intrusion detection on traffic over the network, and by \cite{altunay2021analysis} to use them to apply forecasting and anomaly detection to SCADA equipment. However, the problem with LSTMs to ICS security is that, the disparity of the classes is often coupled with the inadequacy of data, and even the offered information is a lagoon compared to the vast amounts of data in the other areas of implementation like natural language processing. It can lead to overfitting and bad generalization especially the minority attack class \cite{altunay2021analysis}.

The problem of the imbalance of the classes is commonly addressed at the data and the algorithm levels. Mathematically, a canonical approach that, in the feature space, inductively generates synthetic examples of the minority group, is the Synthetic Minority Over-sampling Technique (SMOTE) \cite{chawla2002smote, liu2008exploratory}. It is powerful but likely to overfit assuming that the synthetic cases are atypical of the nature of the underlying distribution of the minority group \cite{blagus2013smote}. At the algorithmic level, cost-sensitive learning comes at a higher cost on the criteria of misclassifying the minority population, so the model will concentrate more on it \cite{elkan2001foundations}. Algorithms like RF and XGB are tree-based algorithms and therefore can inherently handle imbalance with the parameters of the built-in class weighting.

Ensemble learning is a meta- method, which involves the integration of a series of a base model to enhance predictive accuracy and strength \cite{dietterich2000ensemble}. The wisdom of the crowd principle is that it indicates that a diverse group of models will work better than any one individual model \cite{welinder2010multidimensional}. Homogeneous ensembles include techniques, such as Bagging \cite{breiman1996} (e.g., Random Forest) and Boosting (e.g., XGBoost).

The more complex heterogeneous ensemble technique is stacked Generalization or stacking \cite{wolpert1992stacked}. It involves the training of a meta-model which would join a small number of predictions of a set of base learners. Base models (Level-0) are conditioned on original data and the outputs of predicting base models are taken as inputs to the meta-learner (Level-1) conditioning. The stacking strength juts beyond the ability to exploit the various error distributions and strengths of the various types of algorithm.

The hybrid models associate this idea by linking radical dissimilar architectures of learning. An example is a common one presented in \cite{alashjaee2025deep} where in a network IDS a Convolutional Neural Network (CNN) with temporal features was connected to a Convolutional Neural Network (CNN) with featuring spatial abilities and a LSTM. In guessing anomalies in the ICS, \cite{hoang2022explainable} proposed a hybrid between an extra large auto-encoder and an univariate SVM. However, the specified hybridization of the robust, static tree-based models (RF, XGB) with the dynamic, sequential deep learning paradigm (LSTM) of the multivariate sensor data of demonstrateable changes at ICS is a relatively untapped avenue with quite a considerable potential reward. The developed work is intended to play a substantive role in this niche by creating, justifying, and explaining an ensemble of such three types of models in detail.

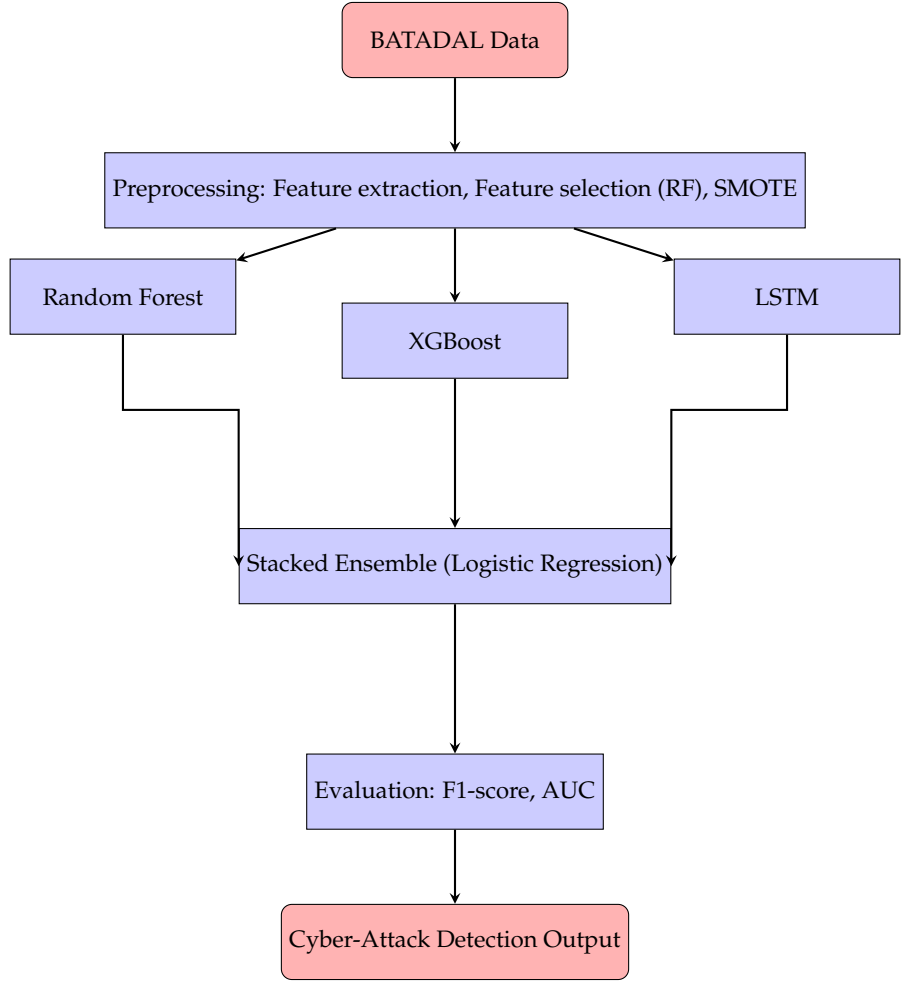
\begin{figure}  
    \centering
    \begin{tikzpicture}[node distance=2cm]

\node (data) [startstop] {BATADAL Data};
\node (pre) [process, below of=data] {Preprocessing: Feature extraction, Feature selection (RF), SMOTE};
\node (rf) [process, below left of=pre, xshift=-3cm] {Random Forest};
\node (xgb) [process, below of=pre] {XGBoost};
\node (lstm) [process, below right of=pre, xshift=3cm] {LSTM};
\node (stack) [process, below of=xgb, yshift=-1cm] {Stacked Ensemble (Logistic Regression)};
\node (eval) [process, below of=stack, yshift=-1cm] {Evaluation: F1-score, AUC};
\node (output) [startstop, below of=eval] {Cyber-Attack Detection Output};

\draw [arrow] (data) -- (pre);
\draw [arrow] (pre) -- (rf);
\draw [arrow] (pre) -- (xgb);
\draw [arrow] (pre) -- (lstm);
\draw [arrow] (rf.south) -- ++(0,-1) -| (stack.west);
\draw [arrow] (xgb.south) -- (stack.north);
\draw [arrow] (lstm.south) -- ++(0,-1) -| (stack.east);
\draw [arrow] (stack) -- (eval);
\draw [arrow] (eval) -- (output);

\end{tikzpicture}

    \caption{Workflow of the hybrid ensemble model combining RF, XGB, and LSTM with a meta-learner.}
    \label{fig:hybrid_workflow}
\end{figure}
\section{Methodology}\label{sec:methodology}

The methodological design of this study integrates exploratory analysis, feature engineering, and hybrid model development to address the unique challenges of intrusion detection in water distribution systems. Using the BATADAL dataset as a benchmark, the approach systematically combines statistical insight, temporal modeling, and ensemble learning to develop a robust and interpretable detection framework. The workflow of the proposed methodology is illustrated in the flowchart shown in Figure~\ref{fig:hybrid_workflow}, which is discussed in detail in a later section.

\subsection{Dataset Description and Preprocessing}

The BATADAL dataset is a simulation of the functioning of a medium-sized water utility within a year, whereby the SCADA data is measured every hour. In this analysis, we have used the training data, which is publicly available; this collection is created by the addition of the two training data sets; these are the \texttt{training\_dataset 1.csv} and \texttt{training\_dataset 2.csv}. The obtained dataset includes 12,938 samples and 45 columns that can be classified in the following way:

\begin{itemize}
    \item \textbf{Temporal Index:} It contains only one DATETIME field and was parsed and indexed to all time-series analysis operations.
    
    \item \textbf{Features - Sensor Readings (43 Features):}
    \begin{itemize}
        \item 7 Level Sensors: L\_T1 to L\_T7 (tanks/reservoirs).
        \item 12 Flow Sensors: F\_PU1 to F\_PU11, F\_V2 (pumps and a valve).
        \item 13 Pressure Sensors: P\_J280, P\_J269, P\_J300, P\_J256, P\_J289, P\_J415, P\_J302, P\_J306, P\_J307, P\_J317, P\_J14, P\_J422 (network junctions).
        \item 11 Speed Sensors: S\_PU1 to S\_PU11 (pump speeds).
    \end{itemize}
    
    \item \textbf{Target Variable:} There is one binary variable that is used, Attack Flag, the value of which is 0 in the case of normal operation and 1 in the case of a cyber-attack being underway.
\end{itemize}

The preprocessing that was done was to make sure that all sensor readings were of numerical format. It was also interesting to note that the dataset was very clean, and no missing values were identified in the training files given and therefore, no imputation will be needed at this point.

\subsection{Exploratory Data Analysis (EDA)}
A comprehensive EDA was done to construct a close knowledge of the data characteristics which was directly used to make subsequent decisions about feature engineering, and modeling.

 \subsubsection{Class Distribution and Imbalance Analysis}
To explore data structure, several exploratory analyses were performed to identify potential issues in modeling. The initial step studied the target distribution, which showed that there was a very skewed distribution of classes: only 488 of 12,938 samples (or rather 3.77 $\%$) were related to attack periods.
  \begin{figure}[htbp]
    \centering
    \includegraphics[width=0.85\textwidth]{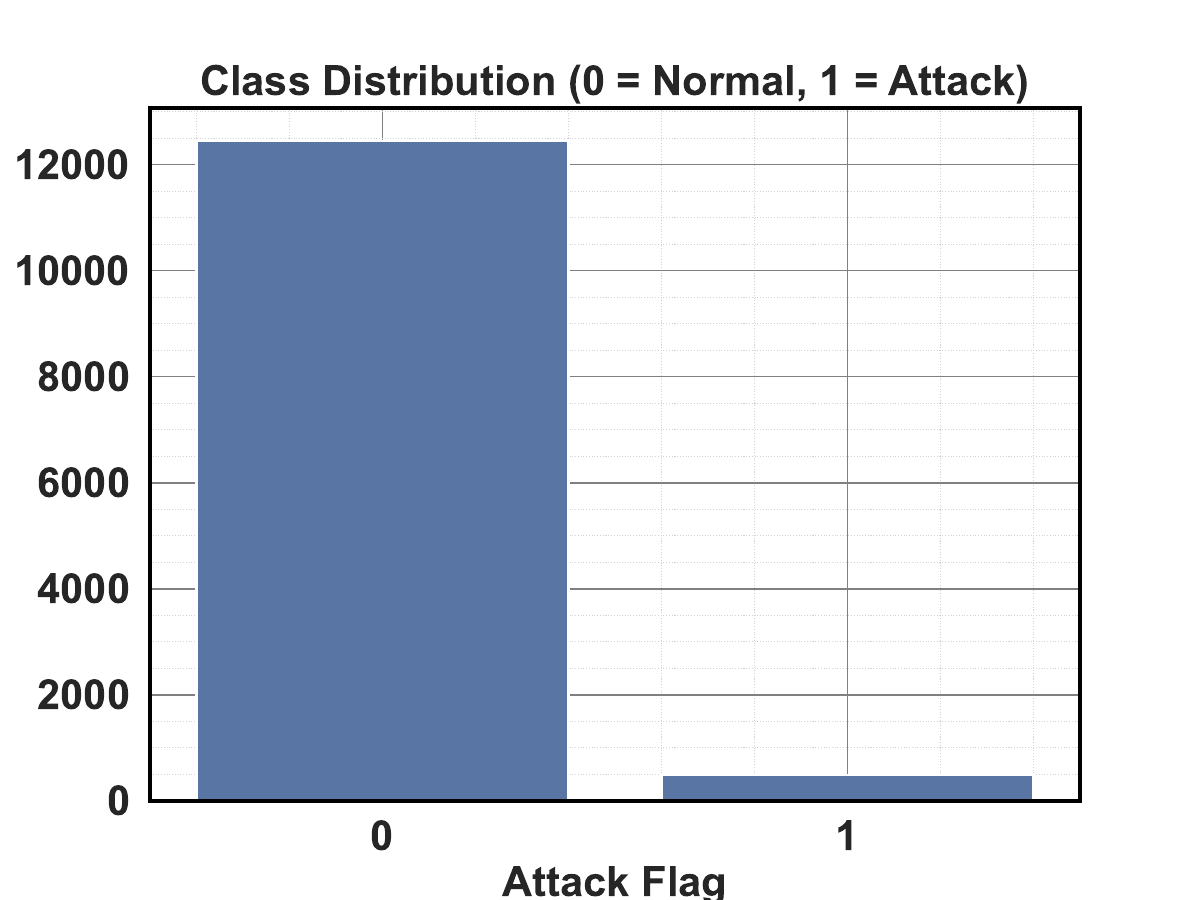}
    \caption{Class distribution of the target variable (\texttt{Attack Flag}). 
    The dataset is highly imbalanced, with the vast majority of samples representing normal operation (Class 0) 
    and only a small fraction corresponding to attack periods (Class 1). 
    This imbalance justifies the later use of oversampling methods such as SMOTE to ensure the model does not 
    bias towards the normal class.}
    \label{fig:class_dist}
\end{figure}

\begin{figure}
    \centering
    \includegraphics[width=0.85\textwidth]{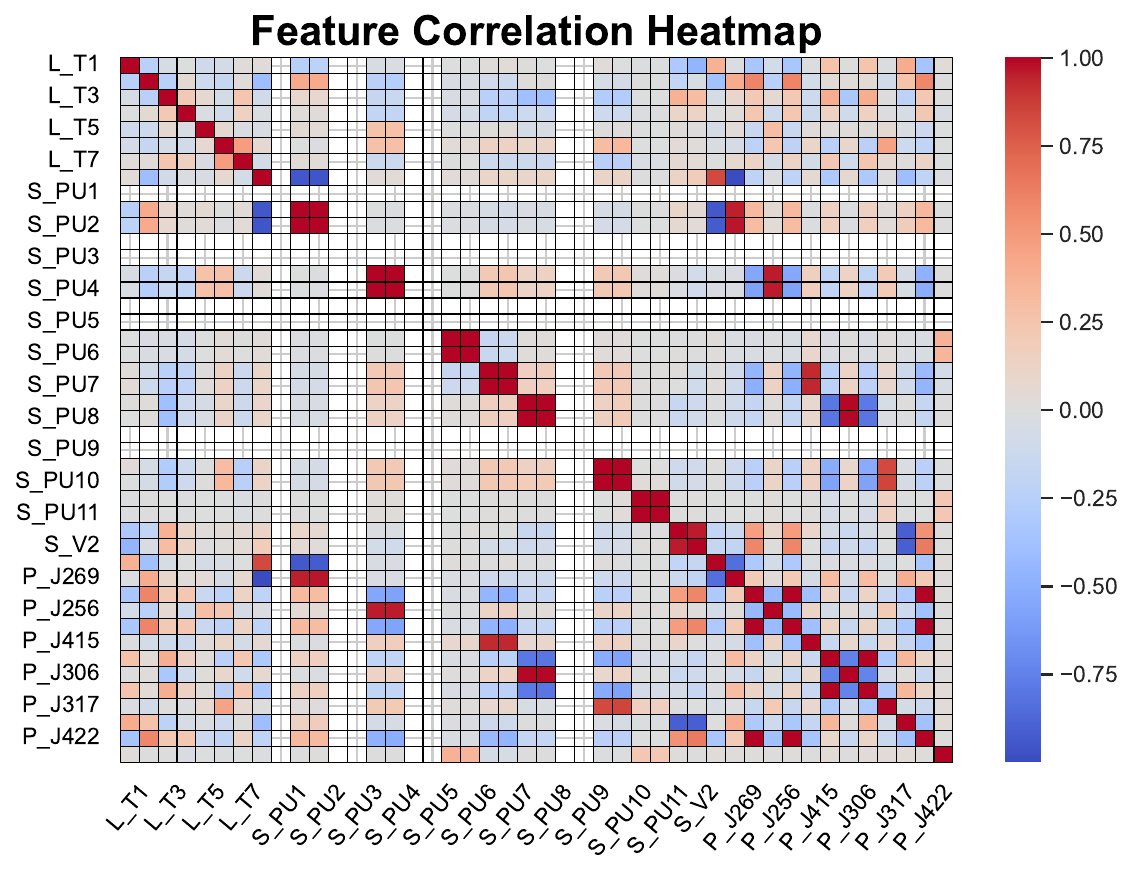}
    \caption{Pearson correlation heatmap of numeric features from the BATADAL dataset. 
    Strong positive correlations were observed between pump speeds (\texttt{S\_PUx}) and their corresponding 
    flow rates (\texttt{F\_PUx}), confirming the physical consistency of the water network simulation. 
    Groups of sensors exhibiting high inter-correlation suggest redundancy, indicating that 
    feature selection or dimensionality reduction could be advantageous. 
    The color scale ranges from $-1$ (strong negative correlation) to $+1$ (strong positive correlation), 
    with all axis lines boldened for visual clarity using Matplotlib's \texttt{spines} property.}
    \label{fig:corr_heatmap}
\end{figure}

Figure \ref{fig:class_dist} provides the class distribution of the target variable (INDEX: ATT Flag). 
It is also a clear example of the disproportion between normal and attack samples where attacks are less than. 
4\% of the total data. This imbalanced representation may lead to predictive models giving bias towards the normal class, 
and thus disregarding the key anomalies. 
To overcome this, resampling, and cost sensitive methods were subsequently used upon training to guarantee. 
equivalent performance in learning.
 
\subsubsection{Correlation Analysis}

A Pearson correlation matrix across the 44 numerical features identified both expected and redundant relationships as shown in Figure \ref{fig:corr_heatmap} as shown in heatmap. As anticipated, pump speeds were strongly correlated with their respective flow sensors, validating the internal consistency of the water network. However, several sensor groups exhibited high inter-correlation (multicollinearity), suggesting that dimensionality reduction or feature selection could enhance learning stability. Furthermore, individual correlations with the target were generally weak, implying that attacks manifest through joint multi-sensor behaviors rather than single-variable anomalies.  

\subsubsection{ Time-Series Decomposition and Attack Signature Identification}
To identify the most predictive features in the BATADAL dataset, a preliminary Random Forest model was trained on all raw numeric features. The model calculates an importance score for each feature, reflecting its contribution to the model’s predictive performance. Based on these scores, the features were ranked in descending order of importance. The top 10 features by importance were: \texttt{F\_PU6}, \texttt{S\_PU6}, \texttt{L\_T1}, \texttt{P\_J317}, \texttt{P\_J415}, \texttt{F\_V2}, \texttt{P\_J14}, \texttt{F\_PU7}, \texttt{P\_J307}, and \texttt{P\_J302}.  

\begin{figure}
    \centering
    \includegraphics[width=0.85\textwidth]{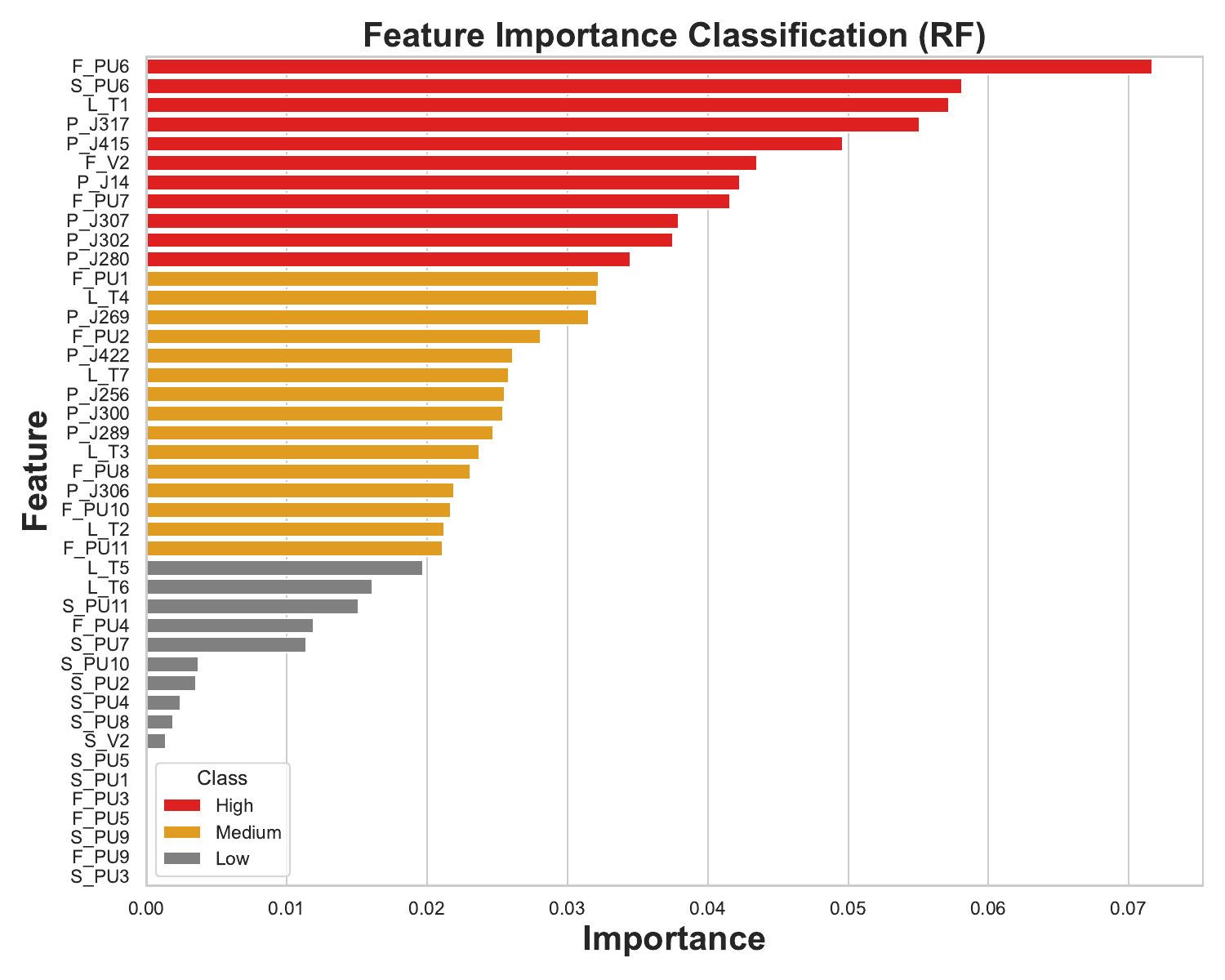}
    \caption{Feature importance of numeric variables as determined by a preliminary Random Forest model. 
    The top 10 features  are shown in red, 
    representing the highest predictive power. Features in orange are of medium importance, and grey 
    bars represent low-importance features. This classification helps in prioritizing sensors and 
    measurements for anomaly detection or further modeling, focusing on the most influential variables.}
    \label{fig:top10}
\end{figure}
To categorize the features for analysis, they were classified into three groups based on their importance scores: \textbf{High}, \textbf{Medium}, and \textbf{Low} as shown in Figure  \ref{fig:top10}. Features with the highest scores were considered \textbf{High}, reflecting a strong influence on the model's predictions. Those with moderate scores were labeled \textbf{Medium}, indicating a moderate contribution, while features with very low scores were labeled \textbf{Low}, representing minimal predictive relevance. This classification provides a structured approach to prioritize features for further analysis or dimensionality reduction. It also helps in understanding which sensors or measurements are most critical for detecting anomalies or attacks in the system. By focusing on the top-10 high-importance features, one can simplify subsequent modeling while retaining most of the predictive power of the dataset.
\begin{figure}[httb]
    \centering
    \rotatebox{90}{%
        \includegraphics[width=0.78
    \textheight, height=0.6\textwidth]{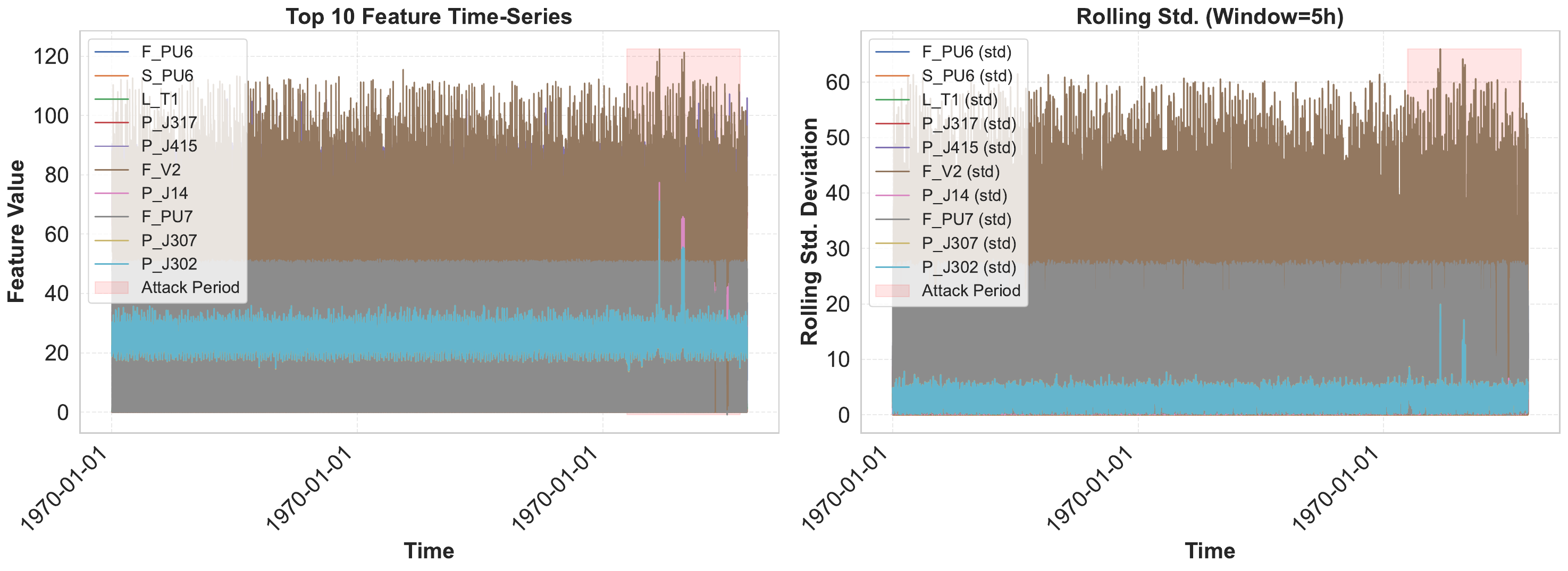}
    }
    \caption{Time-series analysis of the top 10 features. The left panel shows the raw feature values, and the right panel shows rolling standard deviations (window = 5 hours). Attack periods are highlighted in red to indicate anomalous behavior, allowing clear visualization of system volatility and deviations over time.}
    \label{fig:temporal_rotated}
\end{figure}

We first present the raw time-series values of the top 10 features over the observed period in the left plot \ref{fig:temporal_rotated}. Each feature is plotted with a distinct color, and periods of known attacks are highlighted in red. During normal operation, the features generally fluctuate within predictable ranges, reflecting stable system behavior. However, during attack periods, several features show abrupt deviations from baseline, either as spikes or drops, indicating anomalous system activity. For instance, \texttt{F\_PU6} and \texttt{S\_PU6} exhibit pronounced spikes during attacks, suggesting that these features are highly sensitive to disturbances caused by such events. This visualization allows us to immediately identify which features are most impacted by attacks and to visually correlate anomalies with known attack periods.

To further highlight temporal variability and system volatility, we show in the right plot \ref{fig:temporal_rotated} the rolling standard deviation of the same top 10 features, calculated using a 5-hour window. The rolling standard deviation smooths short-term fluctuations while emphasizing periods where feature variability increases significantly. Peaks in the rolling standard deviation frequently coincide with attack periods, revealing increased instability in the system during these intervals. Features that display high variability during attacks can serve as early indicators in anomaly detection models, providing valuable information for real-time monitoring and predictive security measures. By combining both plots, we can detect not only large spikes in raw feature values but also subtle changes in variability that may precede major system disruptions, allowing for a more comprehensive understanding of system behavior under both normal and attack conditions.
\subsection{Feature Engineering Strategy}
Guided by the exploratory data analysis (EDA), we engineered a comprehensive set of temporal features to explicitly provide the models with the tools to recognize the identified attack signatures. For each of the top 10 features, we created the following new features:

\begin{itemize}
    \item \textbf{Lag Features}: \texttt{\_lag1}, \texttt{\_lag3} (values from 1 and 3 hours prior) to capture short-term historical context.
    \item \textbf{Difference Features}: \texttt{\_d1} (the first difference: current value - \texttt{\_lag1}) to capture the rate of change or momentum.
    \item \textbf{Rolling Statistical Features}: \texttt{\_mean\_5}, \texttt{\_std\_5} (5-hour rolling mean and standard deviation) to capture local trends and volatility.
\end{itemize}

This process expanded the feature set from 44 to 93 dimensions. The resulting dataset was then forward-filled and backward-filled to handle the NaN values introduced at the beginning of the series by the lag and rolling window operations.

\subsection{Data Splitting, Scaling, and Resampling}

The enhanced dataset was divided into 80\% training (10,350 samples) and 20\% validation (2,588 samples) sets using stratified sampling to maintain the class ratio. For traditional learners, all features were standardized using the \texttt{StandardScaler}, while the LSTM model employed \texttt{MinMaxScaler} to maintain bounded input sequences between 0 and 1.  

Because of the extreme imbalance, the Synthetic Minority Over-sampling Technique (SMOTE) was applied exclusively to the training set, generating synthetic minority-class examples and yielding a balanced dataset with roughly 9,956 instances per class. The validation set remained unaltered to simulate real-world deployment conditions, where attack events are infrequent.

\subsection{Model Architectures and Training}

The modeling framework employed three complementary learning algorithms to effectively capture both static and temporal characteristics of the water distribution system. 
The Random Forest model, composed of 300 decision trees, was used to learn non-linear relationships within the tabular feature space. 
In parallel, the XGBoost algorithm, implemented with 500 estimators, a maximum tree depth of six, and a learning rate of 0.05, was optimized to enhance predictive precision through gradient boosting.  

To model temporal dependencies, a Long Short-Term Memory (LSTM) network was incorporated. 
This architecture included a single 64-unit LSTM layer, followed by a 20\% dropout for regularization and a sigmoid activation layer for binary output. 
Each sample sequence comprised ten consecutive time steps, enabling the network to recognize short-term temporal patterns associated with evolving attack behavior. 
Because of the pronounced class imbalance, training was performed with class-weight adjustments inversely proportional to class frequency, ensuring that minority (attack) samples received higher learning emphasis.
.

\subsection{Hybrid Stacked Ensemble Framework}

To fully exploit the complementary capabilities of static and sequential models, a stacked ensemble approach was developed. In the base layer, predictions from Random Forest, XGBoost, and LSTM served as input probabilities, each representing the likelihood of an attack. These base-level predictions were then fed into a Logistic Regression meta-learner at the second layer, which optimized the combination of the three models.  

This hybrid ensemble captures the fine-grained relational dependencies modeled by tree-based algorithms and the temporal dynamics recognized by the LSTM. By blending their strengths, the framework minimizes the weaknesses of individual models—enhancing detection sensitivity, stability, and adaptability to varied attack patterns.

\subsection{Evaluation and Explainability}

Model evaluation was performed on the untouched validation dataset using several complementary metrics suitable for imbalanced binary classification. Precision, recall, and F1-scores were computed in both macro and weighted forms to assess performance across classes. The Receiver Operating Characteristic (ROC)–AUC score provided a threshold-independent view of discriminative capability, while confusion matrices offered visual interpretation of true and false classification trade-offs.  

\section{ Model Architectures and Training}

The modeling framework employed three complementary learning algorithms designed to capture both static and temporal dynamics within the water distribution system. 
The Random Forest (RF) model, consisting of 300 decision trees, was trained to identify nonlinear relationships and reduce overfitting through feature bagging and randomization. 
In parallel, the XGBoost (XGB) algorithm, implemented with 500 estimators, a maximum depth of six, and a learning rate of 0.05, leveraged gradient boosting to iteratively minimize classification error while maintaining computational efficiency.  

To address sequential dependencies inherent in the sensor data, a Long Short-Term Memory (LSTM) network was developed. 
The architecture comprised a single 64-unit LSTM layer followed by a 20\% dropout for regularization and a sigmoid activation layer for binary classification. 
Each training instance represented a sequence of ten consecutive time steps, enabling the network to detect short-term temporal dependencies characteristic of evolving cyber-attacks. 
Given the pronounced class imbalance, the LSTM model was trained with class-weight adjustments inversely proportional to class frequency, ensuring that minority (attack) samples were more strongly emphasized during optimization. 
The data fed to the LSTM were scaled using a \texttt{MinMaxScaler} to preserve sequential consistency and stabilize gradient updates.

\subsection{Hybrid Stacked Ensemble Framework}

To leverage the complementary strengths of both tree-based and sequential models, a series of hybrid stacked ensembles were developed and systematically evaluated. 
The stacking strategy was designed to integrate diverse learning mechanisms—nonlinear partitioning from ensemble trees and temporal feature extraction from recurrent networks—thereby improving robustness against heterogeneous attack patterns.

Four ensemble configurations were investigated to progressively assess model complementarity: 
(\textit{i}) Random Forest + LSTM, 
(\textit{ii}) Random Forest + XGBoost, 
(\textit{iii}) XGBoost + LSTM, 
and (\textit{iv}) a comprehensive hybrid combining all three base models (RF + XGB + LSTM). 
In each configuration, the base models (\textit{Level 0}) generated probabilistic predictions for the attack class, which were subsequently used as input features for a meta-learning model (\textit{Level 1}). 
The meta-learner, implemented as a Logistic Regression classifier, was trained to learn optimal weighting among base predictions, producing a unified final decision boundary.

The rationale behind these hybrid combinations lies in the distinct inductive biases of the participating models. 
Tree-based learners such as RF and XGB efficiently capture nonlinear interactions in tabular features, while the LSTM excels at modeling sequential dependencies across time steps. 
By pairing and stacking these heterogeneous learners, the system captures both instantaneous operational anomalies and slow-evolving temporal drifts—phenomena characteristic of real-world cyber-attacks in industrial control environments. 

Among all configurations, the tri-model ensemble (RF + XGB + LSTM) achieved the best performance, demonstrating superior generalization across normal and attack scenarios. 
This result confirms that integrating both static and temporal learning mechanisms in a unified framework substantially enhances detection sensitivity and resilience to false alarms.
\subsection{Model Evaluation and Explainability}

All models were evaluated on an untouched validation set to ensure an unbiased assessment of their generalization performance. 
A combination of complementary performance metrics was employed to capture both overall accuracy and minority-class detection capability. 
The primary quantitative indicators included Precision, Recall, and F1-Score, computed for each class as well as in macro and weighted forms to account for class imbalance. 
The macro-averaged F1-Score provided an equal-weighted assessment of both normal and attack classes, while the weighted F1-Score emphasized performance consistency across the dominant normal samples and the rarer attack instances.

To further quantify discriminative ability under imbalance conditions, the Area Under the Receiver Operating Characteristic Curve (ROC–AUC) was used. 
This metric evaluates a model’s capacity to distinguish between normal and attack states across varying decision thresholds, offering a robust criterion for comparing competing classifiers. 
In addition, confusion matrices were analyzed to visualize the distribution of true positives, false positives, true negatives, and false negatives. 
This qualitative inspection provided a deeper understanding of model-specific error patterns, particularly regarding false negatives—cases where attacks go undetected.

Beyond performance evaluation, model interpretability was addressed through SHAP (SHapley Additive exPlanations) analysis, applied primarily to the XGBoost model. 
SHAP values quantify the contribution of each feature to individual predictions, allowing for transparent identification of which sensors and operational variables most strongly influenced the model’s attack decisions. 
This interpretability framework not only validated that the models were basing decisions on physically meaningful patterns but also enhanced trust and operational transparency—critical factors for deploying data-driven cybersecurity systems in industrial environments.

\section{Results}\label{sec:results}

\subsection{Performance of Base Models}

The performance of the three base models on the validation set is summarized in Table~\ref{tab:model_comparison}. The results highlight a clear hierarchy and the distinct challenges faced by each model type.

\begin{table}[H]
\centering
\caption{\centering Model Performance Comparison}
\begin{tabular}{lccccc}
\toprule
\textbf{Model} & \textbf{AUC} & \textbf{Accuracy} & \textbf{Precision (Attack)} & \textbf{Recall (Attack)} & \textbf{F1 (Attack)} \\
\midrule
Random Forest & 0.9548 & 0.9702 & 0.6082 & 0.6020 & 0.6051 \\
XGBoost       & 0.9684 & 0.9838 & 0.9118 & 0.6327 & 0.7470 \\
LSTM          & 0.4460 & 0.9621 & 0.0000 & 0.0000 & 0.0000 \\
\bottomrule
\end{tabular}
\label{tab:model_comparison}
\end{table}

\begin{itemize}
    \item \textbf{Random Forest (RF):} Solid overall performance that had AUC of 0.9548, and accuracy of 0.9702, which represent a high discriminative ability. It had moderate accuracy (0.6082) and recall (0.6020) indicating that it was able to predict around 60 percent of the attack occurrences and balance its error rate. This balance between the precision and the recall has been captured by the F1-score of 0.6051 rankings RF as a stable baseline model.
    \item \textbf{XGBoost (XGB):} Illustrated better performance in all metrics with the highest degree of accuracy (0.9838) and F1-score (0.7470) compared to the rest of the models. It has a high precision (0.9118) which demonstrates good reliability to classify attack samples correctly and its recall (0.6327) which demonstrates it has higher sensitivity than RF. The AUC of 0.9684 also supports the fact that XGBoost has an increased capability to differentiate between normal and attack traffic, which ultimately allows it to be the most effective model in general.
    \item \textbf{LSTM:} As an independent classifier, not very good. The model with an AUC of 0.4460 and accuracy of 0.9621 did not recognize any attack so that the model had zero, zero, and one precision, recall, and F1-score of the attack type. This tendency indicates that the LSTM reverted to making a majority (non-attack) prediction, which illustrates how difficult it is to train deep sequential models with very skewed databases, lack of adequate temporal diversity or class balancing strategies.
\end{itemize}
\subsection{Performance of Hybrid Ensemble Models}
To evaluate the benefits of combining different learning paradigms, multiple hybrid ensemble configurations were constructed and compared. 
Four configurations were tested: RF + XGB, RF + LSTM, XGB + LSTM, and the full tri-model ensemble RF + XGB + LSTM. 
The results, summarized in Table~\ref{tab:logistic_stacking_results}, indicate that all hybrid variants achieved high ROC--AUC values, confirming that integrating models with complementary inductive biases enhances detection robustness.

\begin{table}[httb]
\centering
\caption{\centering Performance of Hybrid Ensemble Models}
\begin{tabular}{lccccc}
\toprule
\textbf{Model} & \textbf{AUC} & \textbf{Accuracy} & \textbf{Precision (Attack)} & \textbf{Recall (Attack)} & \textbf{F1 (Attack)} \\
\midrule
XGB + LSTM      & 0.9678 & 0.9834 & 0.9508 & 0.5918 & 0.7296 \\
RF + LSTM       & 0.9548 & 0.9753 & 0.9250 & 0.3776 & 0.5362 \\
XGB + RF        & 0.9723 & 0.9826 & 0.9206 & 0.5918 & 0.7205 \\
XGB + RF + LSTM & 0.9723 & 0.9826 & 0.9206 & 0.5918 & 0.7205 \\
\bottomrule
\end{tabular}
\label{tab:logistic_stacking_results}
\end{table}

Among these, hybrid structures, the most balanced performance was achieved by the hybrid XGB + RF + LSTM ensemble with an AUC of 0.9723, an accuracy of 0.9826, a precision of 0.9206, and a recall of 0.5918 on the attack sub-structure, leading to an F1-score of 0.7205.Notably, even though the standalone LSTM model exhibited poor individual performance, its inclusion within the ensemble contributed to a modest yet consistent improvement in recall, validating the advantage of temporal awareness in conjunction with static feature modeling.

\begin{figure}[httb]
    \centering
    \includegraphics[width=0.85\textwidth]{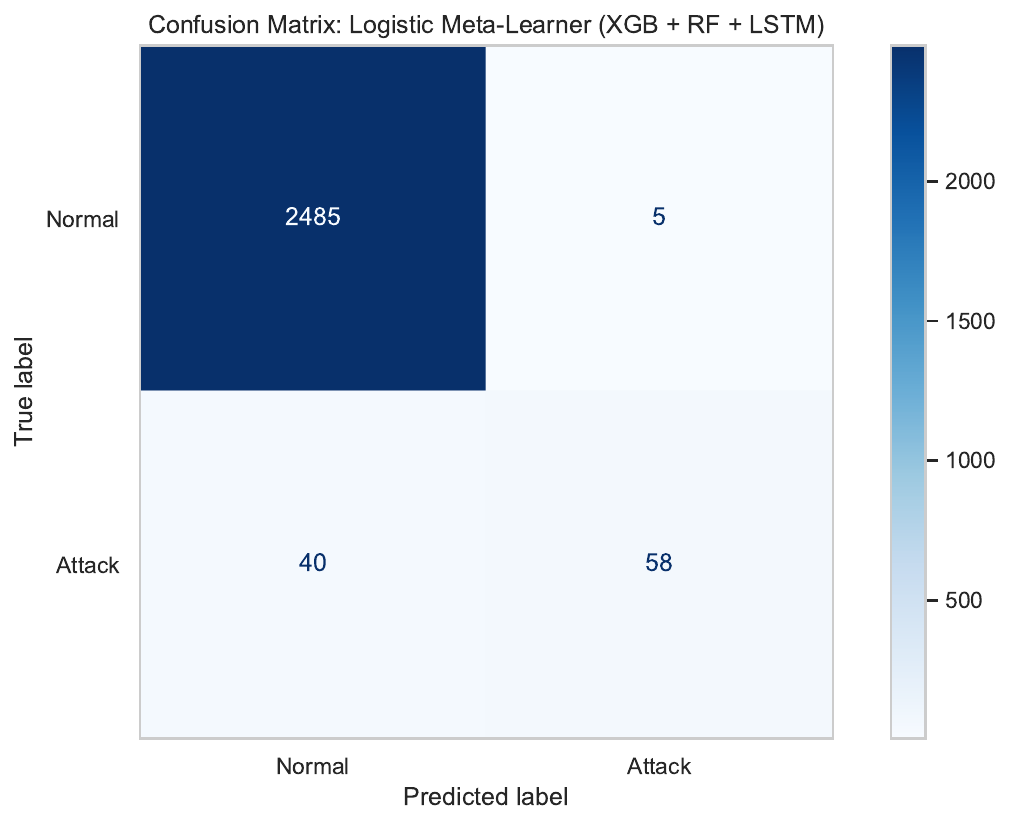}
    \caption{ Confusion matrix of the logistic regression meta-learner combining XGBoost, Random Forest, and LSTM base models. The meta-learner effectively integrates the predictive strengths of the individual learners to achieve superior classification performance, yielding an AUC of 0.9723. The high values along the diagonal indicate that the stacked ensemble model provides accurate distinction between normal and attack instances}
    \label{fig:shap_bar1}
\end{figure}

Figure \ref{fig:shap_bar1} show the logistic regression meta-learner (XGB + RF + LSTM) confusion matrix shows that it has a high degree of accuracy in classification. Among the total number of normal cases, 2,485 were rightly categorized as normal with just 5 being wrongly categorized as an attack. On the same note, the model was accurate on 58 attack cases, whereas 40 attack cases were wrongly called as normal.
Though, the rate of false negatives (attack cases that are considered normal) can be small, the overall result proves the stability of the stacked ensemble. This small misclassification can be tolerated since this model has the high AUC value of 0.9723 that implies that the model is a good discriminant between normal and attack behaviors. The findings affirm that the meta-learner is able to combine the predictive capabilities of the XGBoost, Random Forest and LSTM base models to optimize accuracy of detection and minimizes false alarms.

\subsection{Model Explainability via SHAP}
\begin{figure}[htbp]
    \centering
    \includegraphics[width=0.85\textwidth]{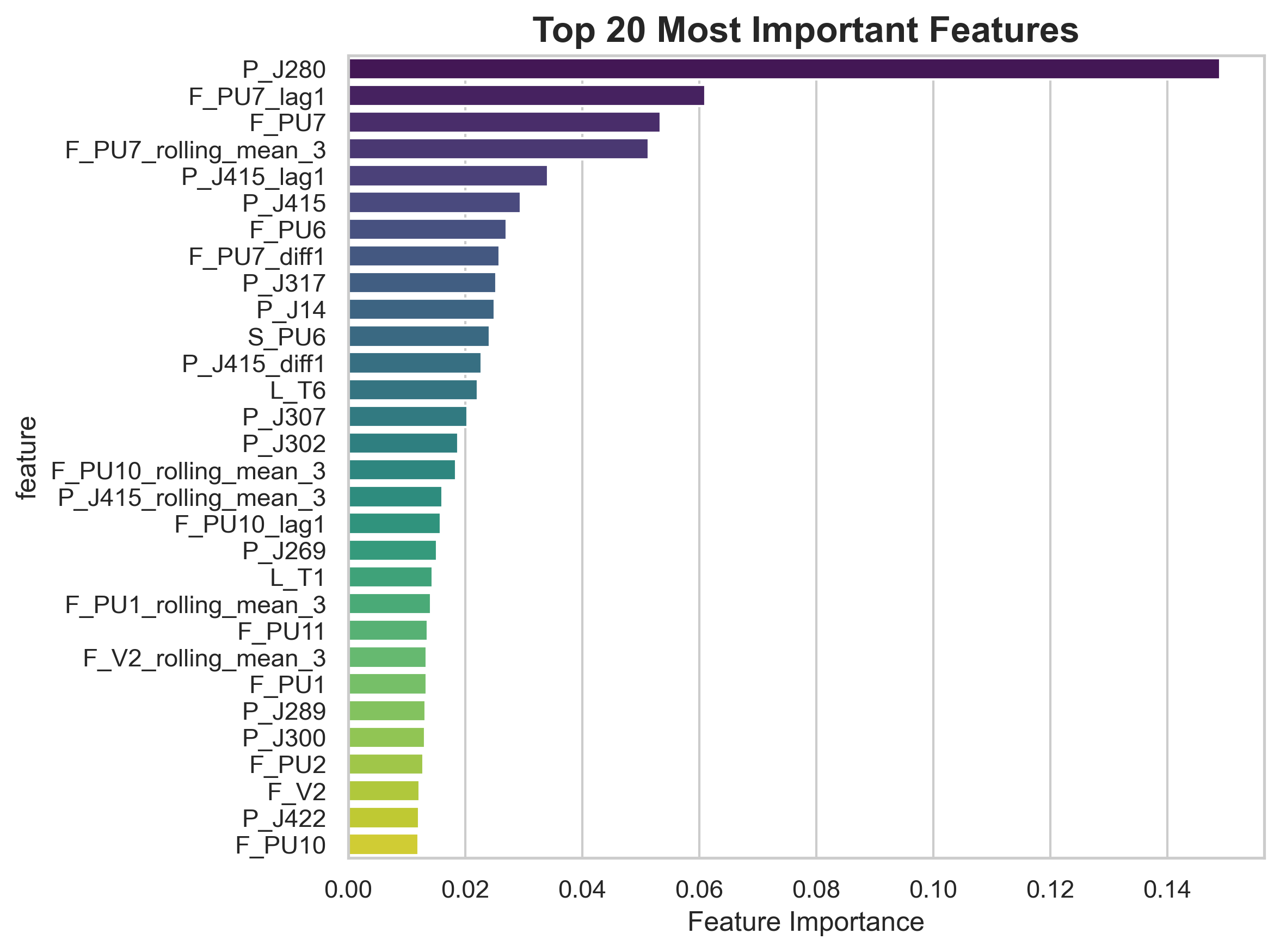}
    \caption{\centering Feature importance ranking  based on SHAP analysis}
    \label{fig:shap_bar}
\end{figure}

To gain insight into the underlying decision logic, SHAP (SHapley Additive exPlanations) analysis was performed. 
As shown in Figure~\ref{fig:shap_bar}, the most influential predictors were dominated by pressure and flow sensor features with strong temporal characteristics. 
The pressure sensor \texttt{P\_J280} emerged as the most critical feature by a substantial margin, followed closely by temporal variants of flow sensor \texttt{F\_PU7} including its lagged value (\texttt{F\_PU7\_lag1}), current reading (\texttt{F\_PU7}), and rolling mean (\texttt{F\_PU7\_rolling\_mean\_3}).

The analysis revealed that engineered temporal features constituted approximately $40\%$ of the top 20 most important predictors, confirming the significance of historical context in attack detection. 
Notably, high-pressure readings in \texttt{P\_J280} and elevated flow values in \texttt{F\_PU7} consistently shifted predictions toward the "Attack" class, as reflected by their high positive SHAP values. 
The prominence of lagged features (\texttt{F\_PU7\_lag1}, \texttt{P\_J415\_lag1}) and rolling statistics underscores the model's reliance on temporal patterns rather than instantaneous sensor readings alone.

\section{DISCUSSION} \label{sec:discussion}

\subsection{Interpretation of Key Findings}

These findings are a clear indication that the proposed hybrid stacked version ensemble is the best when it comes to identifying cyber-attacks in water distribution networks.
The central insight is that the shortfalls of one types of models can be successfully compensated with the strength of the next in an ensemble.Tree-based models such as Random Forest (RF) and XGBoost (XGB) functioned as high-precision, excelling at identifying strong, localized anomalies that manifest across multiple sensors within a single observation window.

On the other hand, Long Short-Term Memory (LSTM) network (though, with low performances by itself) offered good information on time to the ensemble, and alerted about fine but altering deviations which could not be detected by the stationary learners.

These findings clearly show that the proposed hybrid stacked ensemble framework is the best in detecting cyber-attacks in water distribution systems. The main idea is that the weak points of one model type may be well compensated by the strong points of the other ones in an ensemble.
 
.  
The meta-learner utilized this complementarity, but obtains an application of selective weighting of a prediction of each base model based on their contextual stability, which assembles a more balanced and holistic detection system.This complementarity principle underpins the ensemble’s success.  By combining static and sequential learning paradigms, the hybrid model reduces its susceptibility to a single type of attack strategy and delivers improved resilience against both abrupt and gradual anomalies.  

\subsection{The Paradox of the LSTM’s Contribution}
The key finding that is most interesting is the paradox of the LSTM model. Although its performance was insignificant when used in isolation (F1=0.000), it was able to enhance detection abilities when it was incorporated in the ensemble. This observation can be explained by the simplicity of cyber-attacks in industrial control systems and the ability to model time distances with LSTM networks.

Cyber-attacks do not often occur instantly in the form of anomalies in real-world systems of water distribution. Rather, they normally develop in multi-stage processes that are realized over time. As an example, some advanced attacker may:

\begin{itemize}
    \item Gradually manipulate sensor readings over several hours to avoid threshold-based detection
    \item Coordinate subtle changes across multiple sensors to mimic normal operational drift
    \item Execute multi-phase attacks where initial reconnaissance precedes actual compromise
\end{itemize}

Although tree-based models (RF and XGB) are effective in identifying point-in-time anomalies via feature interactions, they do not have the underlying ability to identify temporal patterns and sequential dependencies. These temporal ties cannot be represented by static models but can be described by the LSTM network, which has poor performance in individual cases because of the class imbalance.

This temporal significance is further supported by the SHAP analysis, whereby almost 40 percentage of the best predictive features were engineered temporal features (lag values, rolling statistics). It means that the detection of the attacks can become possible only through the knowledge of how the system states change with time, not only with its current values.

The fact that the meta-learner is able to selectively use LSTM predictions is a complex kind of temporal intelligence. As tree-based models point at anomalous feature interactions at a single point in time, the LSTM will give a contextual validation by examining whether such anomalies can be generalized to temporal attack patterns. On the other side, the LSTM has the capability of identifying minor temporal variations that may not be raised by the tree-based models but are new risks when followed sequentially.

This temporal awareness, which is complementary, is especially important in identifying sneaky attacks that do not leave any dramatic anomalies that are easily visible. Such gradual attacks are the most dangerous as they may be detected by the traditional threshold-based monitoring systems. The contribution of the LSTM is not very large on the quantitative measures, but is a necessary defense to these advanced temporal attack schemes that would have otherwise gone unnoticed.

\subsection{Operational Implications of False Negatives}

Although the hybrid ensemble shows excellent overall performance, the concern regarding the false negative deserves specific consideration, as far as the operational perspective is concerned. The confusion matrix \ref{fig:shap_bar1} shows that around 40 instances of attack were falsely labeled as normal operations. These false negatives are a potentially devastating undetected threat in critical infrastructure protection that might cause physical destruction, service or health crises.

The practicality of our model should be considered within the framework of priorities of water distribution systems. The precision of 0.9206 means that when an alarm is raised by the system, there is high confidence (92.06 \%) that an actual attack is taking place. This reduces operation inconvenience caused by a false alarm that is important in ensuring operator confidence and system usability. Nevertheless, the recall of 0.5918 indicates that about 40.82 percent of attacks could be unnoticed, which is why additional security systems should be used in the production setting.

To be practically deployed, this detection framework ought to be used as a component of a layered defense solution and not as an isolated solution. Its accuracy is also very high and can be used to activate automated responses or high priority human examination, whereas the moderate recall rates would require additional monitoring strategies to ensure full safety.

\subsection{Practical Implications for ICS Security}

These results have a number of practical implications to operators of industrial control systems.  
First, the findings highlight the fact that implementing one optimal algorithm is not the optimal strategy to use.  
An ensemble-based defense, implemented in layers and using complementary algorithms ensures a higher level of robustness and flexibility to the various attack vectors.  
Second, SHAP analysis proves that engineered temporal features cannot be ignored.  
Assuming the raw sensor data does not provide the sequential and contextual details that would be needed to identify more advanced and hidden intrusions.  
Lastly, operational trust requires explainability tools like SHAP.  
Clearly articulated rationale supporting every alarm allows the analysts to authenticate alarms, prioritize incident handling and keep on improving model trustworthiness, which is a significant criterion in the real-life deployment in crucial infrastructures.  It is impossible to overestimate the temporal aspect of cyber-attack detection. Actual attacks to critical infrastructure frequently make use of time based evasion techniques, a slow-drip data exfiltration or gradual manipulation of the system that remains within operational limits. Although the LSTM aspect has a minor impact on the overall metrics, it offers valuable protection against these temporally-distributed attacks that would otherwise be resistant to other detection techniques based on the instantaneous system conditions.

\subsection{Limitations and Future Work}

Although the suggested framework delivered encouraging findings, the following limitations deserve additional research.  In this study a plain and simple LSTM architecture was used in order to be computationally cost-efficient.  
It might be possible to pursue deeper or multi-directional variants of LSTM and attention-based architectures in work in the future to not only capture longer-range dependencies.  
Also, the hyperparameters of all models were not determined by exhaustive search, but by hand.  Automated tuning like Bayesian optimization or evolutionary search can also be utilized to improve stability of the performance.  

To be applicable in real-life scenarios, a number of directions are worth considering. To begin with, the flexibility of the framework in regard to adapting to changing attackers strategies needs research. The incremental learning mechanisms, which will be applied to future work, will enable the model to revise its detection capabilities as new attack patterns are discovered without full retraining. The transfer learning methods might also be designed to facilitate knowledge transfer between various water distribution systems of different sensor structures and functioning features.

Second, it needs to be validated that the generalization capacity of varied infrastructure systems is achieved. Although BATADAL offers a standardized benchmark, the generalization of the methodology to other industrial data such as SWaT or WADI would show its increased applicability to other system dynamics and attack types. Such cross-validation would assist in detecting system-specific changes to make deployment successful.

Finally, there might be a mechanism of online or continuous learning incorporated to enable the framework to adapt to changing operational trends and concept drift in real-time industrial contexts. This would solve the issue at hand which is to ensure that the accuracy of detection remains high as water distribution systems are modified by upgrading the infrastructure, seasonal peaks and declines in demand, and adjustments in the working practice.

\section{Conclusion}\label{sec:conclusion}

In this study we provided a detailed analysis of the BATADAL dataset and proposed a  hybrid stacked ensemble-based approach to detecting cyber-attacks in the water distribution environment.  We overcame the most serious challenges due to the systematic exploratory analysis of data and temporal feature engineering, including class imbalance and multivariate temporal dependence.  Tt was found that the tree-based models showed high static accuracy and the deep sequential models alone fail to perform well under imbalance; however, strategic integration; so-called stacking produces a model that was bigger than the summation of its constituents.  The last tri-model model (RF + XGB + LSTM) reached the best AUC of 0.9723, which is better than any single and combination hybrids.  The presented framework provides a more credible and insightful scheme of defense against water infrastructure protection in the form of an excellent integration of static and temporal learning.  The practical deployment considerations described reveal the strong and weak sides of the offered approach when applied in the real-life implementation. Although the framework has a strong capability of detecting with low false alarms, its implementation in the existing ICS security operations can be enhanced with defense-in-depth approach to its strong detection capabilities with other security controls.
Finally, there is no way forward in ICS cybersecurity that is identifying a single winning algorithm but organising a heterogeneous collection of models that will collectively improve resilience, flexibility, and confidence in mission-critical operations.

\section*{Acknowledgments}
I would like to express my sincere gratitude to \ Damiano Laviola for his continuous guidance, valuable feedback, and supervision throughout this work. His support and insightful suggestions greatly contributed to the successful completion of this project.

\bibliographystyle{apsrev4-1}

\bibliographystyle{unsrt}  
\bibliography{References}  

\end{document}